\documentclass[11pt]{article}
\usepackage{amsmath}
\usepackage{amssymb}
\usepackage{enumerate}
\usepackage{graphicx}
\usepackage{epsfig}
\usepackage[T1]{fontenc}
\usepackage{theorem}
\usepackage{xcolor}
\usepackage{algorithmic, algorithm}
\newcommand{\Frac}[2]{\displaystyle{\frac{#1}{#2}}} 

\newcommand{\scal}[2]{{\left\langle{{#1}\mid{#2}}\right\rangle}}

\newcommand{\menge}[2]{\big\{{#1}~\big |~{#2}\big\}}

\newcommand{\HH}{\ensuremath{\mathcal H}}
\newcommand{\GG}{\ensuremath{\mathcal G}}

\newcommand{\emp}{\ensuremath{{\varnothing}}}

\newcommand{\argmin}{\ensuremath{\operatorname{argmin}}}

\newcommand{\I}{\ensuremath{\operatorname{I}}\,}
\newcommand{\ID}{\ensuremath{\operatorname{I}_D}\,}
\newcommand{\RR}{\ensuremath{\mathbb{R}}}

\newcommand{\RPP}{\ensuremath{\left]0,+\infty\right[}}

\newcommand{\RX}{\ensuremath{\left]-\infty,+\infty\right]}}
\newcommand{\NN}{\ensuremath{\mathbb N}}

\newcommand{\JJ}{\ensuremath{\mathcal H}}

\newcommand{\ZZ}{\ensuremath{\mathbb Z}}

\newcommand{\pinf}{\ensuremath{{+\infty}}}

\newcommand{\dom}{\ensuremath{\operatorname{dom}}}
\newcommand{\prox}{\ensuremath{\operatorname{prox}}}

\newcommand{\sign}{\ensuremath{\operatorname{sign}}}
\newcommand{\inte}{\ensuremath{\operatorname{int}}}

\newtheorem{theorem}{Theorem}[section]
\newtheorem{proposition}[theorem]{Proposition}

\theoremstyle{plain}{\theorembodyfont{\rmfamily}
}
\newtheorem{assumption}[theorem]{Assumption}
\newtheorem{example}[theorem]{Example}

\title{Relaxing Tight Frame Condition in Parallel Proximal Methods for Signal Restoration \thanks{Part of this work appeared in the conference proceedings of EUSIPCO 2010 \cite{Pustelnik_N_2010_eusipco_pro_mir}. This work was supported by the Agence Nationale de la Recherche under grant ANR-09-EMER-004-03.}}
\author{Nelly Pustelnik\thanks{N. Pustelnik is with the Laboratoire de Physique de l'ENS Lyon, CNRS UMR 5672, F69007 Lyon, France. Phone: +33 4 72 72 86 49, E-mail: \texttt{nelly.pustelnik@ens-lyon.fr}.},  Jean-Christophe Pesquet, \\and Caroline Chaux.
  \thanks{J.-C. Pesquet and C. Chaux are with the Universit{\'e} Paris-Est, LIGM, CNRS-UMR 8049, 77454 Marne-la-Vall{\'e}e
Cedex 2, France. Phone: +33 1 60 95 77 39, E-mail: \texttt{\{jean-christophe.pesquet,caroline.chaux\}@univ-paris-est.fr}.}
}

\begin{document}

\maketitle

\begin{abstract}
A fruitful approach for solving signal deconvolution problems consists
of resorting to a frame-based convex variational formulation. In this context,
parallel proximal algorithms and related alternating direction methods
of multipliers have become popular optimization techniques
to approximate iteratively the desired solution. Until now, in most of
these methods,  either Lipschitz differentiability properties or
tight frame representations were assumed. In this paper, it is shown that it
is possible to relax these assumptions by considering a class of non
necessarily tight frame representations, thus offering the
possibility of addressing a broader class of signal restoration
problems. In particular, it is possible to use
non necessarily maximally decimated filter banks with perfect
reconstruction, which are common tools in digital signal processing.
The proposed approach allows us to solve both frame analysis and
frame synthesis problems for various noise distributions. In our simulations, it is applied to the deconvolution of data corrupted with
Poisson noise or Laplacian noise by using (non-tight) discrete dual-tree wavelet representations and filter bank structures.
\end{abstract}


\newpage
\section{Introduction}
Many works in signal/image processing are concerned with 
data restoration problems.
 For such problems, the original data $\overline{y}\in\boldsymbol{\ell}^2(\ZZ)$ is
 degraded by  a stable convolutive operator $T \colon \boldsymbol{\ell}^2(\ZZ) \to
 \boldsymbol{\ell}^2(\ZZ)$  and by a non-necessarily additive
 noise.\footnote{$\boldsymbol{\ell}^2(\ZZ)$ denotes the space of 
discrete-time real-valued signals defined on $\ZZ$ having a finite energy.} The resulting
 observation
model can be written as 
$z = \mathcal{D}_{\alpha}(T \overline{y})$ 
where $\mathcal{D}_{\alpha}$ denotes the noise effect and $\alpha>0$
is some related parameter (for example, $\alpha$ may represent the
variance for Gaussian noise or the scaling parameter for Poisson
noise). In this context, our objective is to recover a signal $y\in
\boldsymbol{\ell}^2(\ZZ)$, the closest possible to $\overline{y}$, from the observation vector
$z$ assumed to belong to $\boldsymbol{\ell}^2(\ZZ)$ and available prior information (sparsity, positivity,\ldots). In early works, this
problem was solved, mainly for Gaussian noise, by using Wiener
filtering, or equivalently quadratic regularization techniques. Later,
multiresolution analyses were used for denoising by applying a
thresholding to the generated coefficients
\cite{Donoho_D_1995_tit_den_st}. Then, in order to improve the
denoising performance, redundant frame representations were
substituted for wavelet bases \cite{Coifman_R_1995_lns_translation_id,Pesquet_J_1996_tsp_tim_iow}.
In
\cite{Combettes_PL_2005_mms_Signal_rbpfbs,Daubechies_I_2004_cpamath_iterative_talipsc,Figueiredo_M_2003_tosp_EM_afwbir,Bect_J_2004_eccv_unified_vfir},
authors considered convex optimization techniques to jointly address
the effects of a noise and of a linear degradation within a
convex variational framework. When the noise is Gaussian,
the forward-backward (FB) algorithm
\cite{Combettes_PL_2005_mms_Signal_rbpfbs} (also known as thresholded
Landweber algorithm when the regularization term is an $\boldsymbol{\ell}_1$-norm
\cite{Daubechies_I_2004_cpamath_iterative_talipsc,Figueiredo_M_2003_tosp_EM_afwbir,Bect_J_2004_eccv_unified_vfir})
and its extensions \cite{Bioucas_J_2007_toip_New_ttsistafir,Beck_A_2009_j-siam-is_fast_istalip}
can be employed in the context of wavelet basis decompositions and its use
can be extended to
arbitrary frame representations
\cite{Chaux_C_2007_ip_variational_ffbip}. However, in the context of a
non-additive noise such as a Poisson noise or a Laplace noise, FB  algorithm is no longer applicable due to the
non-Lipschitz differentiability of the data fidelity term. Other
convex optimization algorithms must be employed such as the Douglas-Rachford (DR) algorithm
\cite{Combettes_PL_2007_istsp_Douglas_rsatncvsr}, the Parallel
ProXimal Algorithm (PPXA)
\cite{Combettes_PL_2008_journal_proximal_apdmfscvip} or the Alternating
Direction Method of Multipliers (ADMM)
\cite{Afonso_M_2009_j-tip_augmented_lacofiip,Setzer_S_2009_Deblurring_pibsbt}. 
These algorithms belong to the class of proximal algorithms and, for
tractability issues, they often require to use tight frame
representation for which closed forms of the involved proximity operators can
be derived \cite{Combettes_PL_2008_journal_proximal_apdmfscvip,Afonso_M_2009_j-tip_augmented_lacofiip,Setzer_S_2009_Deblurring_pibsbt,Weiss_P_2010_sjsc_sol_ctv}.  The goal of this paper is to propose a way to relax the tight frame requirement by considering an appropriate class of frame representations.

In the following, we consider two general convex minimization
problems, which are useful to solve frame-based restoration problems
formulated under a Synthesis Form (SF) or an Analysis Form (AF).
The SF can be expressed as:

\begin{equation}
\label{eq:SA}
\underset{\substack{y = F^{*}x,\; x\in \boldsymbol{\ell}^2(\ZZ)}}{\operatorname{minimize}}\;  \sum_{r=1}^R f_r(L_r y) +  \sum_{s=1}^{S} g_s(x)
\end{equation}

\noindent  and the AF is:

\begin{equation}
\label{eq:AA}
\underset{y\in \boldsymbol{\ell}^2(\ZZ)}{\operatorname{minimize}} \; \sum_{r=1}^R f_r(L_r y) +  \sum_{s=1}^{S} g_s(F y).
\end{equation}

\noindent $F\colon  \boldsymbol{\ell}^2(\ZZ) \to  \boldsymbol{\ell}^2(\ZZ)$ (resp. $F^* \colon  \boldsymbol{\ell}^2(\ZZ) \to  \boldsymbol{\ell}^2(\ZZ)$) denotes the frame analysis (resp. synthesis) operator.  
For every $ r\in\{1,\ldots,R\}$, $f_r\colon \boldsymbol{\ell}^2(\ZZ)\to \RX$ is a
convex, lower semicontinuous, and proper function, $L_r \colon
\boldsymbol{\ell}^2(\ZZ) \to  \boldsymbol{\ell}^2(\ZZ)$ is a stable convolutive operator, and  for every $ s\in\{1,\ldots,S\}$,
$g_s\colon \boldsymbol{\ell}^2(\ZZ) \to \RX$ is a convex, lower semicontinuous, and proper function. 
In several works, SF has been preferred since AF appears to be more difficult to
solve numerically
\cite{Elad_M_2007_j-ip_analysis_vssp,Chaar_L_2009_spie_solving_ipotcot,Carlavan_M_2010_j-trait-signal_reg_ppi,Selesnick_I_2009_spie_signal_rowtcasp}. In
the proposed framework, both approaches have a similar complexity.

This paper is organized as follows: in Section~\ref{sec:frames},
the class of frames considered in this
work is defined 
and their connections with filter bank structures is emphasized. 
In Section~\ref{sec:prox_algo}, we show how these (non necessarily tight) 
frames can be combined
with parallel proximal algorithms in order to solve Problems \eqref{eq:SA}
and \eqref{eq:AA}. The proposed approach is also applicable to related
augmented Lagrangian approaches.
 Finally, restoration results
are provided in Section~\ref{sec:results} for scenarios involving Poisson
noise or Laplace noise by using Dual-Tree Transforms (DTT) and filter bank representations.

\noindent\textbf{Notation:} Throughout this paper, $\Gamma_0(\HH)$ designates the class of lower semicontinuous convex 
functions $\varphi$ defined on a real Hilbert space $\HH$
and
taking their values in $\RX$, which are proper in the sense that 
their domain
$
\dom \varphi=\menge{u\in\HH}{\varphi(u)<\pinf} 
$
is nonempty.
If $\varphi\in\Gamma_0(\HH)$ has a unique 
minimizer, it is denoted by $\underset{u\in\HH}{\argmin} \varphi(u)$.
The interior of a set $C$ is denoted by $\inte C$.

\section{Frame representations}
\label{sec:frames}

\subsection{Definitions}
Physical properties of the target signal
$\overline{y}\in \boldsymbol{\ell}^2(\ZZ)$, such as sparsity or spatial regularity,
may be suitably expressed in terms of its coefficients $\overline{x} = (\overline{x}(k))_{k\in \ZZ}\in \boldsymbol{\ell}^2(\ZZ)$
where $\overline{y}=\sum_{k=-\infty}^\pinf \overline{x}(k) e_k$ and $(e_k)_{k\in \ZZ}$ denotes a 
dictionary of signals in $\boldsymbol{\ell}^2(\ZZ)$.  
Such a dictionary constitutes a frame if there exist two constants $\underline{\mu}$ and $\overline{\mu}$ 
in $\RPP$ such that

\begin{equation}
\label{e:frame1}
\big(\forall y\in\boldsymbol{\ell}^2(\ZZ)\big)\qquad\underline{\mu}\|y\|^2\leq\sum_{k=-\infty}^\pinf|\scal{y}{e_k}|^2\leq \overline{\mu}\|y\|^2.
\end{equation}
The associated frame operator is the injective linear operator defined as

\begin{equation}
\label{e:defF}
\big(\forall y \in \boldsymbol{\ell}^2(\ZZ)\big)\qquad
Fy = \big(\scal{y}{e_k}\big)_{k\in \ZZ},
\end{equation}
the adjoint of which is the surjective linear operator given by

\begin{equation}
\big(\forall x=\big(x(k)\big)_{k\in \ZZ} \in \boldsymbol{\ell}^2(\ZZ)\big)\qquad
F^*x =\sum_{k=-\infty}^\pinf x(k)\, e_k.
\end{equation}
When $F^{-1}=F^*$, an orthonormal basis is obtained.
Further constructions as well as a detailed account 
of frame theory in Hilbert spaces can be found in
\cite{Han_D_2000_book_frames_bgr}.

A tight
frame is such that, for some $\mu \in \RPP$, $F^*F = \mu \I$ 
where $\I$ denotes the identity operator. In
Condition~\eqref{e:frame1}, this means that $\underline{\mu} =
\overline{\mu} = \mu$. 
A simple example of a tight frame is the union of $\mu$ orthonormal 
bases. Other examples of tight frames can be found in 
\cite{Candes_EJ_2002_as_Recovering_eiipipoocf,Daubechies_I_1992_book_ten_lw,Do_MN_2005_tip_coutourlet_tctaedmir}.

\subsection{A class of non necessarily tight frames}
\label{ssec:NTF}

We consider a linear operator $F$ which is basically obtained by
cascading a non necessarily maximally decimated filter bank and a
semi-orthogonal transform. A linear operator $U\colon
\big(\boldsymbol{\ell}^2(\ZZ)\big)^N \rightarrow\big(\boldsymbol{\ell}^2(\ZZ)\big)^Q$, with $N\in \NN^*$ and $Q\in \NN^*$, is said to be
  semi-orthogonal if there exists $\mu_U \in \RPP$
such that $U^* U = \mu_U \I$. Recall that an analysis filter bank can
be put under its polyphase form by performing a polyphase decomposition
$\Pi_D$ followed by a real MIMO (Multi-Input Multi-Output) filtering
$V$ \cite{Vaidyanathan_P_1993_book_mul_sfb}:\\
\noindent \textbullet \;\;the polyphase decomposition $\Pi_D$ is an operator from
  $\boldsymbol{\ell}^2(\ZZ)$ to $(\boldsymbol{\ell}^2(\ZZ)\big)^{D}$ with $D\in \NN^*$ 
such that, for every
$y = \big(y(n)\big)_{n\in \ZZ}\in \boldsymbol{\ell}^2(\ZZ)$, 
$\Pi_D y = (y^{(j)})_{1\leq j\leq D}$ where $y^{(j)} = \big(y(Dn-j+1)\big)_{n\in
  \ZZ}$ is the $j$-th polyphase component of order $D$ of the signal
$y$.
The adjoint operator of $\Pi_D$ is given by
 $\Pi_D^*\colon (\boldsymbol{\ell}^2(\ZZ)\big)^{D} \to \boldsymbol{\ell}^2(\ZZ)\colon
(y^{(j)})_{1\le j \le D} \mapsto u=\big(u(n)\big)_{n \in \ZZ}$
where, for every $n\in \ZZ$ and $j\in \{1,\ldots,D\}$,
$u(Dn-j+1) = y^{(j)}(n)$. So, $\Pi_D^*$ allows us to concatenate $D$
square summable sequences into a single one. It can be noticed that
$\Pi_D^* \Pi_D = \I$ and $\Pi_D \Pi_D^* = \I$, which means that
$\Pi_D$ is an isometry and $\Pi_D^{-1} = \Pi_D^*$.\\
\noindent \textbullet \;\;The MIMO filter $V$ is defined as

{\small{ \begin{equation}
V = \begin{bmatrix}
V_{1,1}&\ldots&V_{1,D}\\
\vdots &      & \vdots\\
V_{N,1}&\ldots&V_{N,D}\\ 
 \end{bmatrix}
\end{equation}}}
where, for every $i\in \{1,\ldots,N\}$ and $j \in \{1,\ldots,D\}$,
$V_{i,j}\colon \boldsymbol{\ell}^2(\ZZ)\to \boldsymbol{\ell}^2(\ZZ)$ is a SISO (Single-Input
Single-Output) stable filter. Hence, the impulse response of this
filter belongs to $\boldsymbol{\ell}^1(\ZZ)$ and its frequency response 
$\widehat{v}_{i,j}$ is a
continuous function. In addition, it is assumed that $V$ is left
invertible, that is: for every $\nu \in [-1/2,1/2]$, the rank of the matrix
$\widehat{\boldsymbol{v}}(\nu) = [\widehat{v}_{i,j}(\nu)]_{1\le i \le N,1\le j \le D}$ is
equal to $D$. The adjoint operator of $V$ is the $D\times N$ MIMO filter given by
{\small{ \begin{equation}
V^* = \begin{bmatrix}
V_{1,1}^*&\ldots&V_{N,1}^*\\
\vdots &      & \vdots\\
V_{1,D}^*&\ldots&V_{N,D}^*\\ 
 \end{bmatrix}
\end{equation}}}
where, for every $i\in \{1,\ldots,N\}$ and $j \in \{1,\ldots,D\}$,
$V_{i,j}^*$ is the SISO filter with complex conjugate frequency response $\widehat{v}_{i,j}^*$.\\
We have then the following result (the proof is provided in Appendix~\ref{ap:F}):

\begin{proposition}\label{prop:F}
The operator $F = \Pi_Q^* U V \Pi_D$ is a frame operator with frame constants
$\underline{\mu} =$\linebreak $ \inf_{\nu\in[-1/2,1/2]} \sigma_{\rm min}(\nu)$
and $\overline{\mu} = \sup_{\nu\in[-1/2,1/2]} \sigma_{\rm
  max}(\nu)$,
where, for every $\nu$, $\sigma_{\rm min}(\nu)\in \RPP$ and $\sigma_{\rm max}(\nu)\in \RPP$ are the
minimum and maximum eigenvalues of $\widehat{\boldsymbol{v}}(\nu)^{\rm H}
\widehat{\boldsymbol{v}}(\nu)$.\footnote{The notation $\widehat{\boldsymbol{v}}(\nu)^{\rm H}$ corresponds to the transconjugate of $\widehat{\boldsymbol{v}}(\nu)$.} 
In addition, we have:
\begin{equation}\label{e:FsF}
 F^* F = \mu_U \Pi_D^* V^* V \Pi_D.
\end{equation}
\end{proposition}

 The resulting frame is not necessarily tight. However, when 
$(\forall \nu \in [-1/2,1/2])$
 $\widehat{v}(\nu)^{\rm
   H}\widehat{v}(\nu) = \ID$, the frame is tight. This includes paraunitary systems $V$ as particular cases
when $D=N$.

Below, we provide examples of popular transforms in signal
processing which belong to the class of considered frame representations.

\subsubsection{Example 1 -- Dual-tree transforms}
\label{sss:dtt}
In order to obtain low redundancy representations, frames such as
the 2D  $M$-band DTT have been proposed
\cite{Kingsbury_N_2001_jacha_com_wsiafs,Selesnick_I_2005_dsp_dual_tdtcwt,Chaux_C_2006_tip_ima_adtmbwt}. The
real (resp. complex) DTT consists of
performing $N=2$ (resp. $N=4$) $M$-band orthonormal wavelet
decompositions in parallel where, for every $i\in\{1,\ldots,N\}$, $U_i
\colon \boldsymbol{\ell}^2(\ZZ)\rightarrow\boldsymbol{\ell}^2(\ZZ)$ denotes the $i$-th
orthogonal wavelet transform.
An orthogonal combination of the subbands modeled by $\Phi \colon
\big(\boldsymbol{\ell}^2(\ZZ)\big)^N\rightarrow \big(\boldsymbol{\ell}^2(\ZZ)\big)^N$ is applied to
ensure directionality properties. $(U_i)_{1\leq i\leq N}$ and $\Phi$ are related to $U$ by the following relation:
{\small{\begin{equation}\label{e:DTT}
U = \Phi\;
\begin{bmatrix}
U_1 &  0 & \ldots & 0\\
0   &  U_2  & \ddots & \vdots\\
\vdots & \ddots & \ddots & 0\\
0   &  \ldots & 0 &  U_N\\ 
\end{bmatrix}.
\end{equation}}}
Since $\Phi^*\Phi = \Phi\Phi^* = \I$ and, for every $i\in \{1,\ldots,N\}$, $U_i^* U_i = U_i U_i^* =
\I$, $U$ is  an orthogonal transform.
In the general form of DTT, each orthonormal wavelet
decomposition is preceded by a prefiltering stage related to the
discretization process. More precisely, it takes the form:
{\small{\begin{equation}
V = \begin{bmatrix}
V_{1,1}^{*} \; \ldots \; V_{N,1}^{*}
\end{bmatrix}^{*}.
\end{equation}}}
The variables $Q$ and $D$ as defined before are thus equal to $N$ and $1$,
respectively. The left invertibility condition
here reduces to the fact that the frequency response 
$|\widehat{v}_{1,1}|^2+\cdots+|\widehat{v}_{N,1}|^2$ does not vanish. 
Note that, due to the presence of prefilters,
the \emph{discrete} DTT is not a tight frame in general.
Furthermore, an extended class of DTTs can be obtained by using tight frame
overcomplete representations $(U_i)_{1\le i \le N}$  having the same frame constant $\mu_U$
(e.g. redundant wavelet representations derived from orthogonal filters
after appropriate renormalization), in which case $U$ as defined
by \eqref{e:DTT} is a semi-orthogonal transform.

\subsubsection{Example 2 -- Filter banks}
\label{sss:fb}
Analysis filter banks with perfect reconstruction correspond to the
case when  $U = \I$ and $Q=N$. $D$ is the decimation factor and 
$N$ is the number of channels.
The redundancy introduced by such a filter bank structure is $N/D \ge
1$. 
More details about filter bank design can be found in
\cite{Strang_G_book_wavelets_fb}. 
Lapped transforms \cite{Malvar_H_book_sig_plt} can also be implemented with filter bank structures.
As already mentioned, filter
banks do not constitute tight frames in general. Note that if $D=1$, a fully
undecimated filter bank is obtained.

Tight frame representations have been widely used 
for data recovery by using convex optimization
methods. In the next section, we recall some
convex optimization tools and show the relevance of the considered class of
frames in recent optimization approaches.

\section{Use of parallel proximal algorithms}
\label{sec:prox_algo}
A number of algorithms such as the alternating split Bregman algorithm \cite{Goldstein_D_2008_splut_bml1rp},
augmented Lagrangian techniques \cite{Afonso_M_2009_j-tip_augmented_lacofiip} and parallel proximal methods \cite{Pesquet_J_2010}
have been recently proposed to address possibly nonsmooth convex
optimization problems encountered in the solution of restoration problems.
Here, we will focus on this specific class of algorithms which
have proven to be useful for solving problems like \eqref{eq:SA} and
\eqref{eq:AA} when a tight frame representation is employed. A common
feature of the aforementioned algorithms is that they require a
large-size linear system to be solved at each iteration. When non
tight frame representations are used, the computational cost of the associated
inversion may become prohibitive. We will see however that for the
class of frames introduced in Section \ref{sec:frames}, this inversion
can be performed in an efficient manner. This fact will be
demonstrated by considering an extension of the Parallel ProXimal
Algorithm (PPXA), hereafter designated as PPXA+. As shown in \cite{Pesquet_J_2010},
PPXA+ is closely related to the alternating direction method of
multipliers and its parallel extensions \cite{Setzer_S_2009_Deblurring_pibsbt}.

We recall that 
the proximity operator \cite{Moreau_J_1965_bsmf_Proximite_eddueh} of a
function $\varphi \in  \Gamma_0(\JJ)$ is defined as
\begin{equation}
 \prox_\varphi\colon\JJ \to \JJ\colon v \mapsto \arg\min_{u \in \JJ} \Frac12\left\|u-v\right\|^{2} + \varphi(u).
\label{eq:prox}
\end{equation}
It can be observed that when $\varphi$ is the indicator function 
$\iota_C$ of a nonempty closed convex subset $C$ of $\JJ$, i.e. it takes on the value $0$ in $C$ and 
$\pinf$ in $\HH \setminus C$, 
$\prox_{\iota_C}$ reduces to the projector $P_C$ onto $C$. Other
examples of proximity operators corresponding to potential
functions of standard log-concave univariate probability densities
have been listed in
\cite{Combettes_PL_2005_mms_Signal_rbpfbs,Chaux_C_2007_ip_variational_ffbip,Combettes_PL_2008_journal_proximal_apdmfscvip}. The
proximity operators employed in the experimental part of this paper, are recalled below.

\begin{example} (soft-thresholding rule) \\
\label{ex:gg}
Let $\alpha>0$, and set $\varphi\colon\RR\to\RR\colon\xi\mapsto \alpha |\xi|$. Then, for every $\xi\in\RR$, $\prox_\varphi\xi=\sign(\xi) \max\{|\xi| - \alpha,0\}$.
\end{example}

\begin{example}(Poisson minus log-likelihood function) {\rm \cite{Chaux_C_2007_ip_variational_ffbip}} \\
\label{ex:gamd}
Let $\alpha>0$, $\chi \ge 0$, and set 

{\small{\begin{align}
\label{e:gamd}
 \varphi\colon&\RR\to\RX \colon \xi\mapsto
\begin{cases}
-\chi\ln(\xi)+\alpha\xi,&\text{if}\;\;\chi>0\;\text{and}\;\xi>0;\\
\alpha\xi,&\text{if}\;\;\chi=0\;\text{and}\;\xi\geq 0; \\
\pinf, & \text{otherwise}.
\end{cases}
\end{align} }}
Then, for every $\xi\in\RR$,
\begin{equation}
\prox_\varphi\xi=\frac{\xi-\alpha+\sqrt{|\xi-\alpha|^2+4\chi}}{2}.
\end{equation}
\end{example}

One of the difficulties in
the resolution of problems like \eqref{eq:SA} or
\eqref{eq:AA} is that they involve linear operators. Unfortunately,
the proximity operator of the composition of a linear operator 
and a convex function takes a closed form expression only under
restrictive assumptions as stated below:

 \begin{proposition}{\rm\cite{Combettes_PL_2007_istsp_Douglas_rsatncvsr}}\\
 \label{p:linprox}
Let $\GG$ be a real Hilbert space, $\varphi\in\Gamma_0(\GG)$  and 
let $L\colon \HH \rightarrow \GG$ denote a bounded linear operator. Suppose that 
 $LL^{*}=\chi\I $, for some $\chi\in\RPP$.
 Then, $\varphi\circ L\in\Gamma_0(\HH)$ and $\prox_{\varphi\circ L}=\I +\chi^{-1}L^{*}(\prox_{\chi \varphi}-\I) L$.
 \end{proposition}
When dealing with linear operators $L$ such that $LL^*\neq \chi \I$
for any $\chi > 0$, proximal algorithms requiring the inversion of a
linear operator
at each iteration can however be designed.
For example, Algorithm~\ref{algo:SA} (resp. Algorithm~\ref{algo:AA})
can be applied to
Problem~\eqref{eq:SA} (resp. Problem~\eqref{eq:AA}).
(In these algorithms, the sequences 
$(a_{r,\ell})_{1\le r \le R}$ and $(b_{s,\ell})_{1\le s
  \le S}$  model possible numerical errors
in the computation of the proximity operators at iteration $\ell$.)

\begin{algorithm}
\caption{\label{algo:SA}}
\begin{algorithmic}
\STATE Initialization \\
$\left \lfloor \begin{array}{l}
(\eta_r)_{1\le r \le R}\in \RPP^R,(\kappa_s)_{1\le s \le S}\in \RPP^S; \;
(v_{r,0})_{1\leq r\leq R} \in \big(\boldsymbol{\ell}^2(\ZZ)\big)^R,(w_{s,0})_{1\leq s\leq S} \in \big(\boldsymbol{\ell}^2(\ZZ)\big)^S \\
x_0 = \arg \min_{u\in \boldsymbol{\ell}^2(\ZZ)} \sum_{r=1}^R \eta_r {\Vert L_r F^{*} u - v_{r,0}\Vert}^2  +\sum_{s=1}^S \kappa_s {\Vert  u - w_{s,0}\Vert}^2\\
\end{array} \right.$
\STATE For $\ell=0,1,\ldots$ \\
$\left \lfloor \begin{array}{l}
\mbox{For} \;\; r=1,\ldots,R \qquad\qquad p_{r,\ell} = \prox_{f_r/\eta_r}v_{r,\ell} + a_{r,\ell}\\
\mbox{For} \;\; s=1,\ldots,S \qquad\qquad r_{s,\ell} = \prox_{g_s/\kappa_s}w_{s,\ell} + b_{s,\ell}\\
\lambda_{\ell} \in ]0,2[\\
c_{\ell} = \arg \min_{u\in \boldsymbol{\ell}^2(\ZZ)} \sum_{r=1}^R \eta_r {\Vert L_r F^{*}  u - p_{r,\ell}\Vert}^2   + \sum_{s=1}^S \kappa_s {\Vert  u - r_{s,\ell}\Vert}^2\\
\mbox{For} \;\; r=1,\ldots,R \qquad\qquad  v_{r,\ell+1} = v_{r,\ell} + \lambda_\ell \big( L_r F^{*}  (2c_{\ell} - x_{\ell}) - p_{r,\ell}\big)\\
\mbox{For} \;\; s=1,\ldots,S \qquad\qquad  w_{s,\ell+1} = w_{s,\ell} + \lambda_\ell \big( 2c_{\ell} - x_{\ell} - r_{s,\ell}\big)\\
x_{\ell+1} = x_{\ell} + \lambda_{\ell}(c_{\ell} - x_{\ell})\\
\end{array} \right.$
\end{algorithmic}
\end{algorithm}
\begin{algorithm}
\caption{\label{algo:AA}}
\begin{algorithmic}
\STATE Initialization \\
$\left \lfloor \begin{array}{l}
(\eta_r)_{1\le r \le R}\in \RPP^R,(\kappa_s)_{1\le s \le S}\in \RPP^S;\;
(v_{r,0})_{1\leq r\leq R} \in \big(\boldsymbol{\ell}^2(\ZZ)\big)^R,(w_{s,0})_{1\leq s\leq S}
\in \big(\boldsymbol{\ell}^2(\ZZ)\big)^S \\
y_0 = \arg \min_{u\in \boldsymbol{\ell}^2(\ZZ)} \sum_{r=1}^R \eta_r {\Vert L_r u - v_{r,0}\Vert}^2 +\sum_{s=1}^S \kappa_s {\Vert F u - w_{s,0}\Vert}^2\\
\end{array} \right.$
\STATE For $\ell=0,1,\ldots$ \\
$\left \lfloor \begin{array}{l}
\mbox{For} \;\; r=1,\ldots,R \qquad\qquad p_{r,\ell} = \prox_{f_r/\eta_r}v_{r,\ell} + a_{r,\ell}\\
\mbox{For} \;\; s=1,\ldots,S \qquad\qquad r_{s,\ell} = \prox_{g_s/\kappa_s}w_{s,\ell} + b_{s,\ell}\\
\lambda_{\ell} \in ]0,2[\\
c_{\ell} = \arg \min_{u\in \boldsymbol{\ell}^2(\ZZ)} \sum_{r=1}^R \eta_r {\Vert L_r  u - p_{r,\ell}\Vert}^2  + \sum_{s=1}^S \kappa_s {\Vert F u - r_{s,\ell}\Vert}^2\\
\mbox{For} \;\; r=1,\ldots,R \qquad\qquad   v_{r,\ell+1} = v_{r,\ell} + \lambda_{\ell} \big( L_r  (2c_{\ell} - y_{\ell}) - p_{r,\ell}\big)\\
\mbox{For} \;\; s=1,\ldots,S \qquad\qquad   w_{s,\ell+1} = w_{s,\ell} + \lambda_{\ell} \big( F (2c_{\ell} - y_{\ell}) - r_{s,\ell}\big)\\
y_{\ell+1} = y_{\ell} + \lambda_{\ell}(c_{\ell} - y_{\ell})\\
\end{array} \right.$
\end{algorithmic}
\end{algorithm}

The convergence of the sequence $(x_\ell)_{\ell \in \ZZ}$
(resp. $(y_\ell)_{\ell \in \ZZ}$) generated by Algorithm~\ref{algo:SA}
(resp. Algorithm~\ref{algo:AA}) to an optimal solution of the
related optimization problem is guaranteed under the following
technical assumptions (see \cite{Pesquet_J_2010} for more details):

\begin{assumption}
 \begin{enumerate}
  \item[]
  \item $\Big(\bigcap_{r=1}^{R} \inte \dom f_r \circ L_r F^{*}\big)
    \cap \Big(\bigcap_{s=1}^{S-1} \inte \dom g_s\Big) \cap \dom g_S\neq \emp $
  \item [] (resp. $\Big(\bigcap_{r=1}^{R} \inte \dom f_r \circ
    L_r\Big)\cap \Big(\bigcap_{s=1}^{S-1} \inte \dom g_s \circ  F\Big) \cap
    \dom g_S \circ F
    \neq \emp $).
  \item There exists $\underline{\lambda}\in]0,2[$ such that $(\forall \ell \in \NN) \; \underline{\lambda}\leq \lambda_{\ell+1} \leq \lambda_{\ell}$.
  \item $(\forall r \in \{1,\ldots,R\})$ $\sum_{\ell= 0}^\pinf \Vert a_{r,\ell} \Vert < +\infty$ and $(\forall s \in \{1,\ldots,S\})$ $\sum_{\ell= 0}^{\pinf} \Vert b_{s,\ell} \Vert < +\infty$.
 \end{enumerate}
\end{assumption}

Note that quadratic minimizations need to be performed in the
initialization step and in the computation of the intermediate
variable $c_\ell$ at iteration $\ell$. These amount to inverting a
large-size linear operator which is untractable for arbitrary
frames. We will now show that the 
 class of frame operators $F$ introduced in Section \ref{ssec:NTF} allows us
 to overcome this difficulty. 

To do so, recall that the convolutive operators $(L_r)_{1\le r \le R}$ can be put
under a polyphase form \cite{Vaidyanathan_P_1993_book_mul_sfb}:
\begin{equation}
(\forall r \in \{1,\ldots,R\})\qquad L_r = W_r \Pi_D
\end{equation}
where the polyphase decomposition operator $\Pi_D$ has been defined in Section~\ref{ssec:NTF} and $W_r = [W_{r,1},\ldots,W_{r,D}]$ is a MISO (Multi-Input Single-Output) filter (for every $j\in \{1,\ldots,D\}$, $W_{r,j}\colon \boldsymbol{\ell}^2(\ZZ)\to \boldsymbol{\ell}^2(\ZZ)$
is a SISO filter). Now, invoking Proposition \ref{prop:F}  
and making use of Sherman-Morrison-Woodbury identity yield:\\
\noindent \textbullet \;\; for SF (Algorithm~\ref{algo:SA})
\begin{align}
\label{eq:sol1}
\kappa \Big(\sum_{r=1}^R  \eta_r F L_r^{*} L_r F^{*}  +
 \kappa \I\Big)^{-1}
 =&  \I - F\big(\sum_{r=1}^R \eta_r L_r^{*} L_r\big)\Big(\kappa
\I + F^{*}F\big(\sum_{r=1}^R \eta_r L_r^{*} L_r\big) \Big)^{-1}F^{*}\nonumber\\
=& \I- F\Pi_D^*\big(\sum_{r=1}^R \eta_r W_r^{*}
W_r\big)\Pi_D\Big(\kappa \I + \mu_U \Pi_{D}^*V^*V \big(\sum_{r=1}^R \eta_r W_r^{*} W_r\big)\Pi_{D}\Big)^{-1}F^{*}\nonumber\\
=& \I- F\Pi_D^*\big(\sum_{r=1}^R \eta_r W_r^{*}
W_r\big) \Big(\kappa \I + \mu_U V^*V\big(\sum_{r=1}^R \eta_r W_r^{*} W_r\big)\Big)^{-1}\Pi_{D}F^{*},
\end{align}

\noindent \textbullet \;\; for AF (Algorithm~\ref{algo:AA})
\begin{align}
\label{eq:sol2}
\Big( \sum_{r=1}^R \eta_r L_r^{*} L_r + \kappa
F^{*}F\Big)^{-1} 
&=\Pi_{D}^*\Big( \sum_{r=1}^R \eta_r W_r^{*} W_r + \kappa \mu_U V^*V\Big)^{-1}\Pi_{D},
\end{align}
where $\kappa = \sum_{s=1}^S \kappa_s$. Note that the idea of using the Woodbury matrix identity to handle the inversion for SF was proposed in \cite{Figueiredo_M_2010_j-tip_res_pia} in the context of tight frames. 
The inversions in \eqref{eq:sol1} (resp. \eqref{eq:sol2}) 
can be performed by noticing that, as $(W_r)_{1\le r \le R}$ and $V$ are
multivariate filters with frequency responses $(\widehat{\boldsymbol
  w}_r)_{1\le r \le R}$ and $\widehat{\boldsymbol v}$, 
$\kappa \I + \mu_U V^*V\big(\sum_{r=1}^R \eta_r W_r^{*} W_r\big)$
(resp. $\sum_{r=1}^R \eta_r W_r^{*} W_r + \kappa \mu_U V^*V$) is a
MIMO filter with frequency response: for every $\nu \in [-1/2,1/2]$,
\begin{equation}
\kappa \ID + \mu_U \widehat{\boldsymbol v}(\nu)^{\rm H}
\widehat{\boldsymbol v}(\nu)\big(\sum_{r=1}^R \eta_r
\widehat{\boldsymbol w}_r(\nu)^{\rm H} \widehat{\boldsymbol
  w}_r(\nu)\big) \qquad \Big(\text{resp.}\; \sum_{r=1}^R \eta_r \widehat{\boldsymbol w}_r(\nu)^{\rm H} \widehat{\boldsymbol
  w}_r(\nu) + \kappa \mu_U \widehat{\boldsymbol v}(\nu)^{\rm H}
\widehat{\boldsymbol v}(\nu)\Big).
\end{equation}
Hence, by resorting to Fast Fourier Transform implementations, the problem
reduces to the inversion of $D\times D$ matrices.

\section{Experimental results}
\label{sec:results}

We apply the proposed optimization method to the restoration of images 
degraded by a blur and a
Poisson (resp. Laplace) noise.  For simplicity sake, the proposed
formulation has been presented in the 1D case, but its 2D extension is
straightforward. 
The scaling factor of the Poisson
(resp. Laplace) noise is $\alpha =0.1$ (resp. $\alpha=10$). The data
fidelity term is related to the Poisson (resp. Laplace)
minus log-likelihood, which corresponds to the potential function 
$\varphi$ defined by \eqref{e:gamd} (resp. by Example~\ref{ex:gg}). 
In the considered examples, the regularization
term simply is an $\boldsymbol{\ell}_1$-norm modeling the sparsity of $\overline{y}$
through a frame representation. Two different frames are
considered: a (Complex) Dual-Tree Transform, presented in
Section~\ref{sss:dtt}, and an eigenfilter bank decomposition
\cite{Patil_B_2008_tcas1_eig_ado, Tkacenko_A_2003_tcas2_eig_dma}, which is a special case of the
subband structure presented in Section~\ref{sss:fb}. 
We constrain the data values to belong to $[0,255]$, so defining a
closed convex constraint set $C$. The considered SF (resp. AF) problem
is a
particular case of Problem~\eqref{eq:SA} (resp. \eqref{eq:AA}) where
$R=2$, $S=1$, $f_1 = \varphi$, $f_2=\iota_C$, and $g_1 = \tau \Vert \cdot
\Vert_1$ with $\tau>0$. The last function corresponds to the
regularization term operating in the frame domain. In the considered
problems, $L_1 = T$ and $L_2 = \I$. The proximity operators associated
to $f_1$, $f_2$, and $g_1$ are derived from Example~\ref{ex:gamd}, the
projection onto $C$, and Example~\ref{ex:gg}. In our simulations, the
parameter $\tau$ is empirically chosen to maximize the
signal-to-noise-ratio (SNR). In general, the value of this parameter
is not the same for FA and FS.

Note that an alternative to the proposed approach consists of
resorting to
primal-dual algorithms
\cite{Chen_G_1994_j-mp_pro_bdm,Chambolle_A_2010_first_opdacpai,Esser_E_2010_j-siam-is_gen_fcf,Combettes_P_2011_j-jmaa_pro_scf,Briceno_L_2011_mon_ssm}.
These algorithms are appealing as they do not require any operator inversion
and they can thus be employed with arbitrary frames.
However,
after appropriate choices for the weights $(\eta_r)_{1\le r \le R}$,
$(\kappa_s)_{1\le s \le S}$, and the relaxation parameter
$\lambda_{\ell}$, the proposed approach appeared to be faster.
For example, similar frequency domain implementations
of the Monotone + Skew Forward Backward Forward (M+SFBF)
algorithm \cite{Briceno_L_2011_mon_ssm} were observed to be about twenty times
slower than the proposed method in term of iterations and computation time.

\begin{figure}
\centering
\begin{tabular}{cccc}
\!\!\!\!\!\!\includegraphics[width=3.6cm]{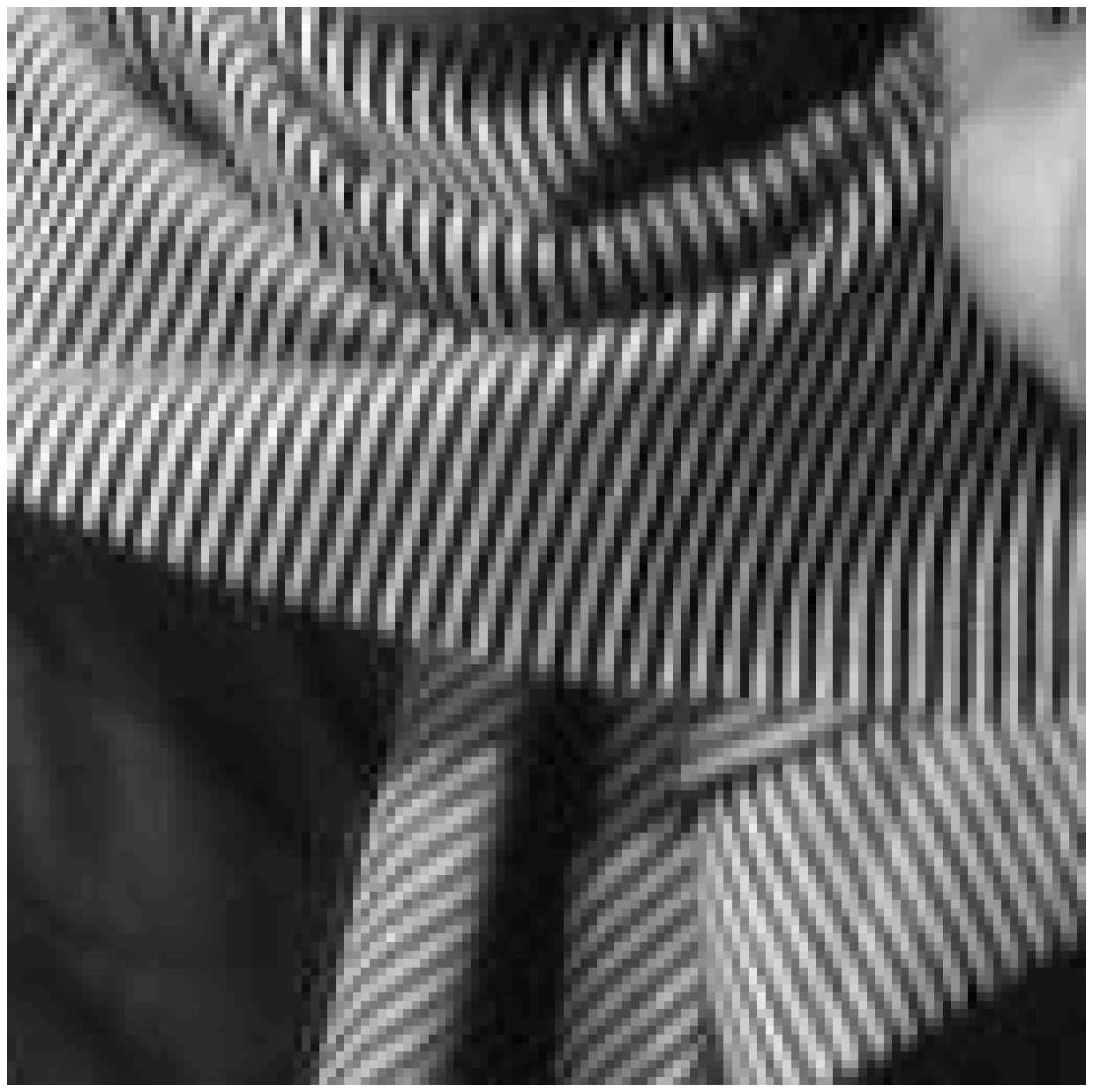}
&
 \includegraphics[width=3.6cm]{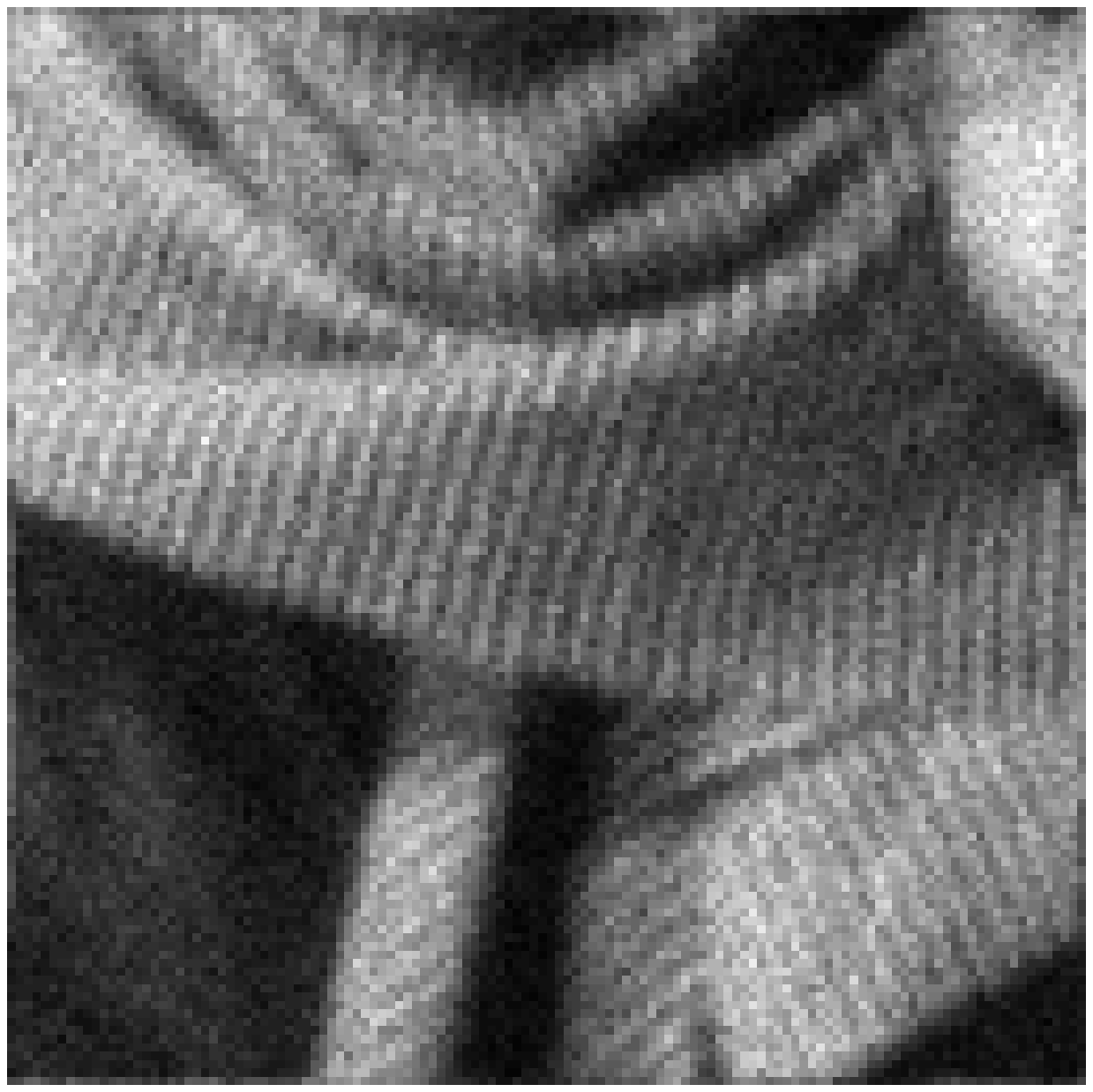}&\includegraphics[width=3.6cm]{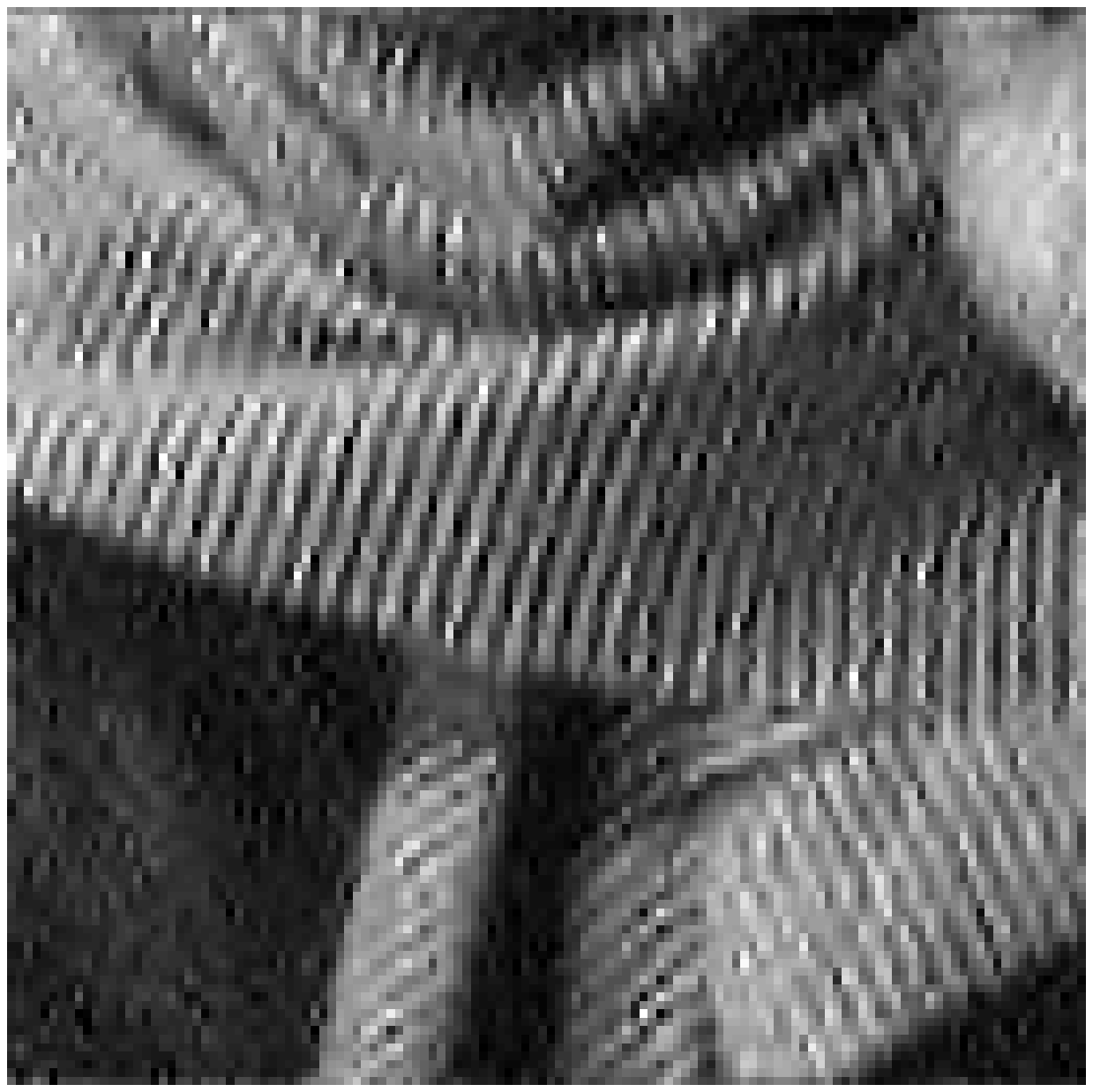} 
&\includegraphics[width=3.6cm]{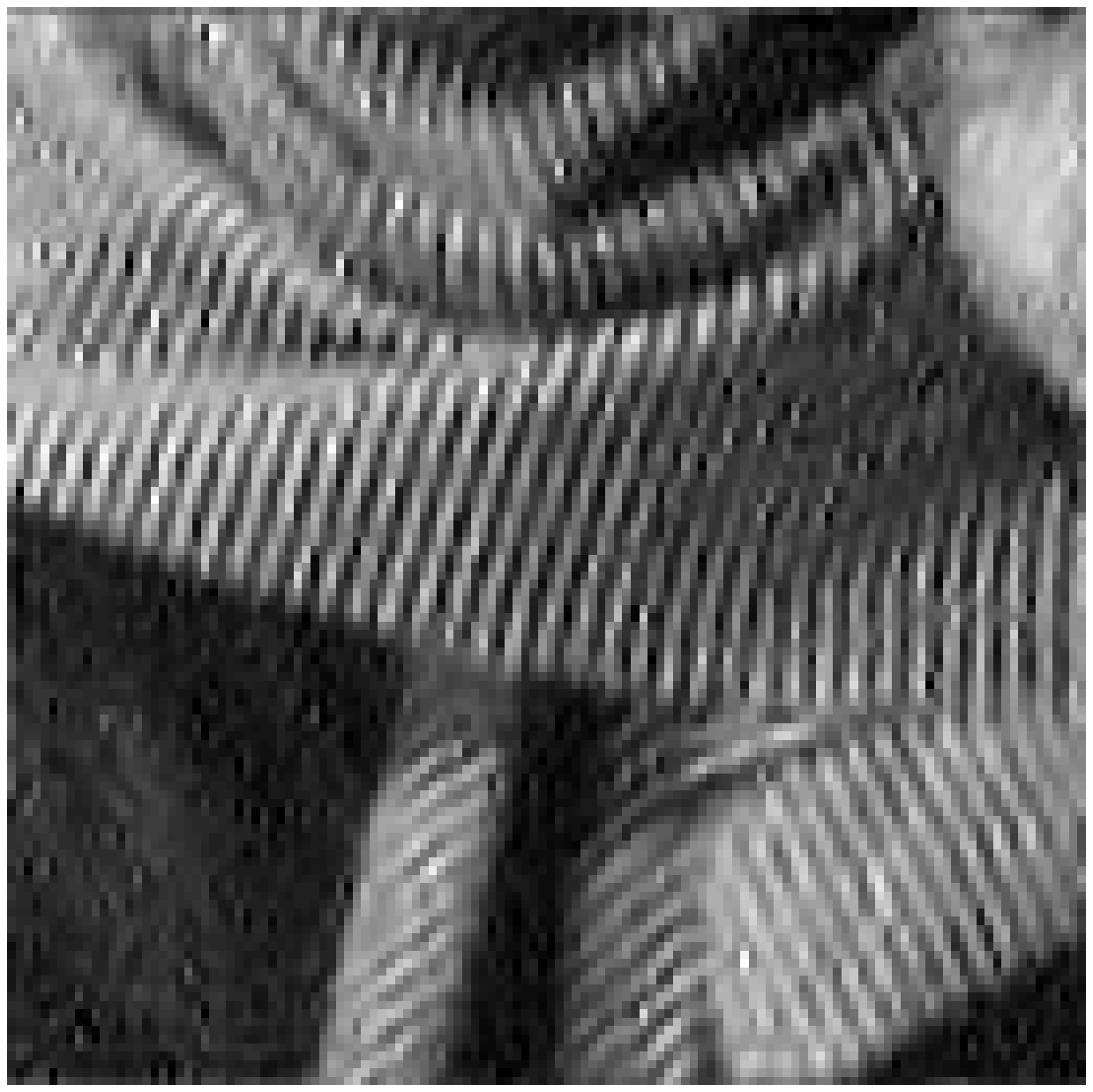}\\
 Original
&
{\small{\!\!\!\!\!\!Degraded}}   & {\small{Tight-frame Complex DTT}} & {\small{Complex DTT}} \\
\!\!\!\!\!\!& {\small{SNR = 11.4 dB}}  & {\small{SNR = 13.3 dB}}   & {\small{SNR = 14.2 dB }} \\
\!\!\!\!\!\!&{\small{SSIM = 0.53}} & {\small{SSIM = 0.69}} & {\small{SSIM = 0.73}}
\end{tabular}
\caption{Cropped versions of Barbara image (size $128\times 128$). Restored images using SF and complex DTT.\label{fig:TF_NTF}}

\end{figure}

\begin{figure}
\centering
\begin{tabular}{cccc}
\!\!\!\!\!\!\includegraphics[width=3.6cm]{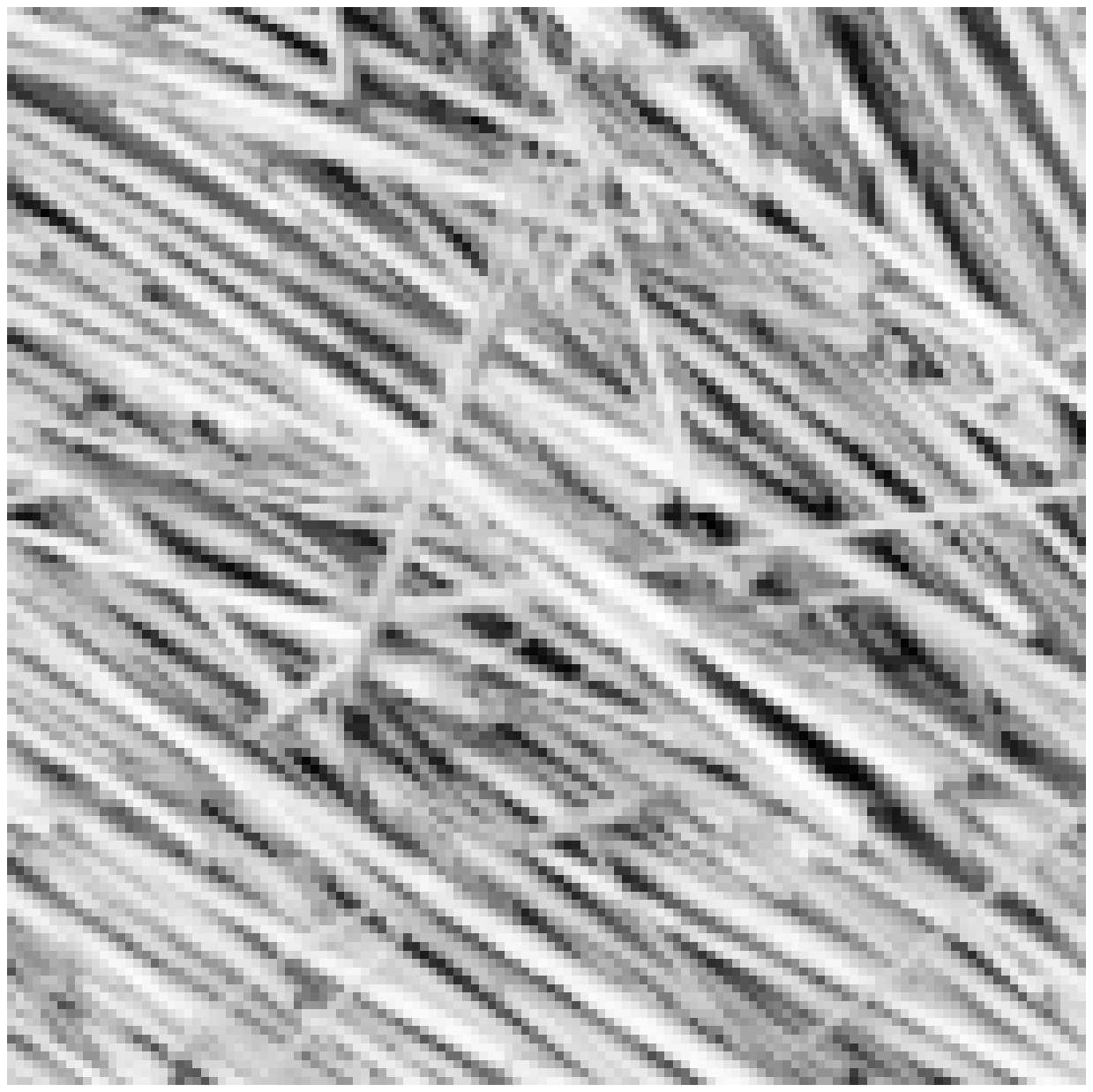}
&
\includegraphics[width=3.6cm]{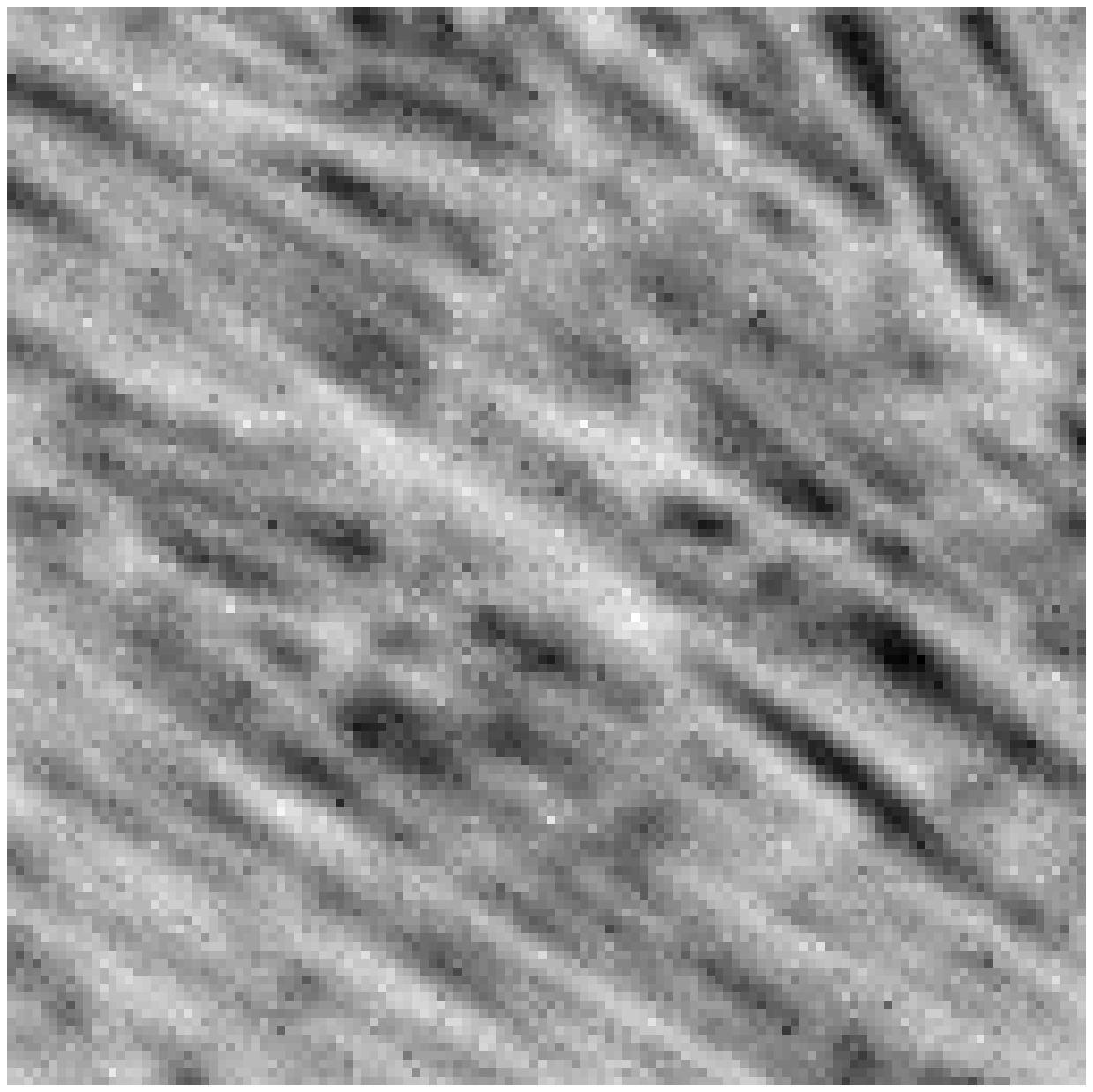} 
&
\includegraphics[width=3.6cm]{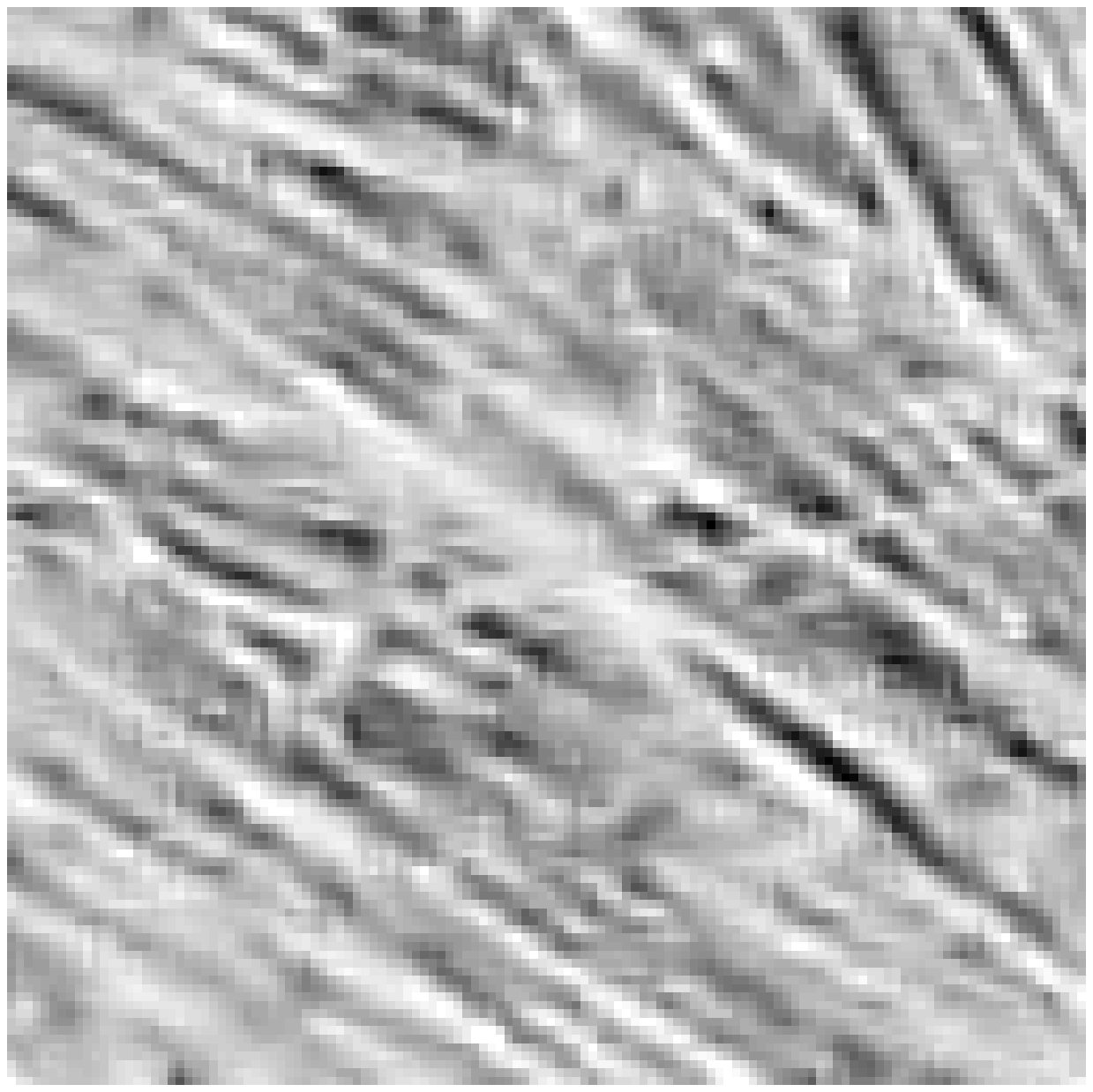}
&
\includegraphics[width=3.6cm]{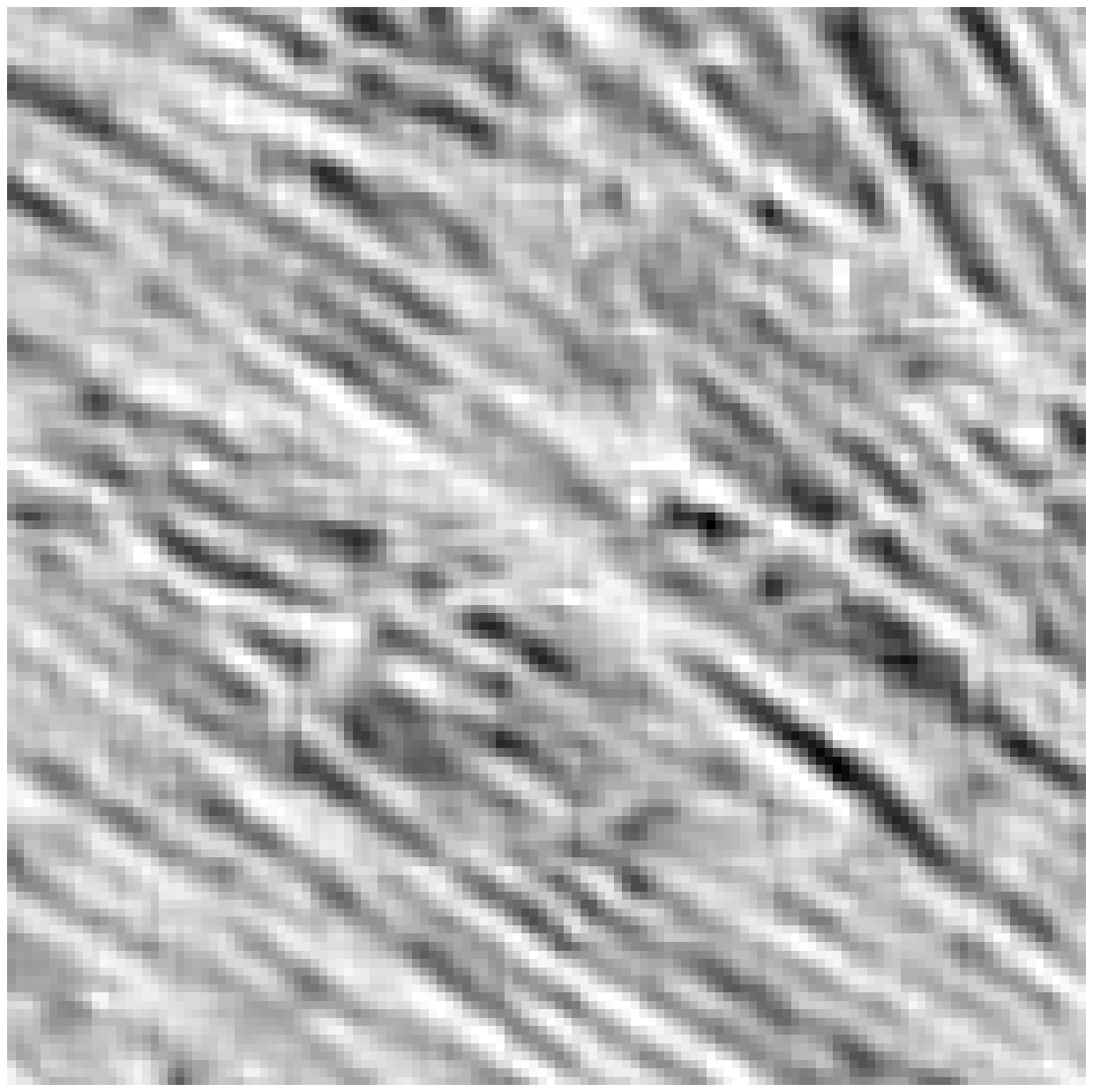} \\
{\small{\!\!\!\!\!\!Original}} & {\small{Degraded}} &  {\small{DTT}} & {\small{Eigenfilter bank}}\\
\!\!\!\!\!\!&{\small{SNR = 14.8 dB}} &  {\small{SNR = 16.7 dB}} & {\small{SNR = 17.3 dB}}\\
\!\!\!\!\!\!&{\small{SSIM = 0.42}}   &{\small{SSIM = 0.64}} & {\small{SSIM = 0.69}}
\end{tabular}
\caption{Cropped versions of Straw image  (size $128\times 128$). Restored images using AF considering DTT and eigenfilter banks.\label{fig:SA_AA}}
\end{figure}

Figure~\ref{fig:TF_NTF} shows the restoration results for a cropped
version of the ``Barbara'' image  in the
presence of Poisson noise and a uniform blur of size $5\times 5$. We
adopt a SF criterion and we consider a tight version (i.e., for
every $i\in\{1,\ldots,4\}$, $V_{i,1} = \I$) as well as a non-tight version of
Complex DTT. The Complex DTT \cite{Chaux_C_2006_tip_ima_adtmbwt} is
computed using symlets of length 6 over 3 resolution levels. In order
to efficiently perform the inversions in \eqref{eq:sol1} and
\eqref{eq:sol2}, fast discrete Fourier diagonalization techniques have
been employed. 
The use of the non-tight Complex DTT including prefilters 
allows us to improve the quality of the
results both visually and in terms of SNR and SSIM \cite{Wang_Z_2009_spm_MES_lioli}. 
Figure~\ref{fig:SA_AA} displays a second restoration example 
for a cropped version of the ``Straw'' image in the presence of Laplace noise and a uniform blur of size $5\times
5$. AF results are presented by using a DTT and 
an eigenfilter bank ($D=8$ and $N=14$) computed from the degraded image. This formulation leads to better results than
those 
obtained with SF. Significant gains in favour of the
eigenfilter bank can be observed .

\section{Appendix}
\subsection{Proof of Proposition \ref{prop:F}} \label{ap:F}

{\small{Eq. \eqref{e:FsF} is a direct consequence of the fact
that $U$ is semi-orthogonal and $\Pi_Q^*$ is an isometry. 
By using Parseval's formula, for every $y\in \boldsymbol{\ell}^2(\ZZ)$,
\begin{equation}
\|Fy\|^2 = \mu_U \|V \Pi_D y\|^2 
= \mu_U\int_{-1/2}^{1/2} \|\widehat{\boldsymbol{v}}(\nu) \widehat{\boldsymbol{y}}(\nu)\|^2 d\nu
\end{equation}

\noindent where $\widehat{\boldsymbol{y}}(\nu) = 
[\widehat{y}^{(1)}(\nu),\ldots,\widehat{y}^{(D)}(\nu)]^\top$,
and, for every $j\in \{1,\ldots,D\}$, $\widehat{y}^{(j)}$
is the Fourier transform of the discrete signal $y^{(j)}$.
By using the fact that, for every $\nu\in [-1/2,1/2]$, $\sigma_{\rm min}(\nu) \|\widehat{\boldsymbol{y}}(\nu)\|^2 \le
\|\widehat{\boldsymbol{v}}(\nu) \widehat{\boldsymbol{y}}(\nu)\|^2
\le \sigma_{\rm max}(\nu) \|\widehat{\boldsymbol{y}}(\nu)\|^2$
we get
\begin{equation}
\mu_U \inf_{\nu\in[-1/2,1/2]} \sigma_{\rm min}(\nu)
\int_{-1/2}^{1/2} \|\widehat{\boldsymbol{y}}(\nu)\|^2 d\nu
\le \|Fy\|^2 \le \mu_U
\sup_{\nu\in[-1/2,1/2]} \sigma_{\rm min}(\nu) \int_{-1/2}^{1/2}
\|\widehat{\boldsymbol{y}}(\nu)\|^2 d\nu.
\end{equation}
Besides, we have
\begin{equation}
\int_{-1/2}^{1/2} \|\widehat{\boldsymbol{y}}(\nu)\|^2 d\nu
= \sum_{j=1}^D \int_{-1/2}^{1/2} |\widehat{y}^{(j)}(\nu)|^2
d\nu =  \sum_{j=0}^{D-1} \sum_{n=-\infty}^{+\infty} |y(Dn-j)|^2
= \sum_{n=-\infty}^{+\infty} |y(n)|^2 = \|y\|^2.
\end{equation}
Now, it can be noticed that since $\widehat{\boldsymbol{v}}$ is
continuous, $\sigma_{\rm min}$ and $\sigma_{\rm max}$ are continuous
functions too. By invoking Weierstrass theorem, there thus exist 
$\underline{\nu} \in [-1/2,1/2]$ and $\overline{\nu}\in [-1/2,1/2]$
such that $\inf_{\nu\in[-1/2,1/2]} \sigma_{\rm min}(\nu) = \sigma_{\rm
  min}(\underline{\nu})$ and $\sup_{\nu\in[-1/2,1/2]} \sigma_{\rm max}(\nu) =
\sigma_{\rm max}(\overline{\nu})$. Finally, $\sigma_{\rm
  min}(\underline{\nu}) > 0$ since 
$\operatorname{rank}\big(\widehat{\boldsymbol{v}}(\underline{\nu})^{\rm
H}
\widehat{\boldsymbol{v}}(\underline{\nu})\big)
= \operatorname{rank}\big(\widehat{\boldsymbol{v}}(\underline{\nu})\big) = D$.

}}
\footnotesize{

}

\end{document}